\documentclass[usenatbib,useAMS]{mn2e}
\usepackage{graphicx}

\usepackage{aas_macros}

\newcommand{\lta}{{\small\raisebox{-0.6ex}
{$\,\stackrel{\raisebox{-.2ex}{$\textstyle <$}}{\sim}\,$}}}

\title[]{Evidence for high accretion-rates
in Weak-Line T Tauri stars?}

\author[]{S.\,P.\ Littlefair$^{1}$, Tim Naylor$^{1}$,
Tim J.\,Harries$^{1}$, Alon Retter$^{2}$ and S.\,O'Toole$^{2,3}$\\
$^1$School of Physics, University of Exeter, Exeter EX4 4QL, UK \\
$^2$School of Physics, University of Sydney, Sydney, NSW 2006, Australia\\
$^3$Universit\"{a}t Erlangen-N\"{u}rnberg, Sternwartstr. 7, D-96049, Bamberg, Germany\\}

\date{\center{\Large Submitted for publication in the Monthly Notices of the
Royal Astronomical Society \\ 
\vspace{.5cm} \today}} 

\begin{document}
\maketitle

\begin{abstract} 
  We have discovered T Tauri stars which show startling spectral
  variability between observations seperated by 20 years. In spectra published
  by \cite{ba92} these objects showed very weak H$\alpha$ emission,
  broad CaII absorption and so called ``composite spectra'', where the
  spectral type inferred from the blue region is earlier than that
  inferred from the red. We present here new spectroscopy which shows
  that all four stars now exhibit strong H$\alpha$ emission, narrow CaII
  emission and a spectral type which is consistent at all wavelengths.
  
  We propose a scheme to understand these changes whereby the
  composite spectra of these stars can be explained by a period of
  active accretion onto the central, young star. In this scheme the
  composite spectrum consists of a contribution from the stellar
  photosphere and a contribution from a hot, optically thick,
  accretion component. The optically thick nature of the accretion
  flow explains the weakness of the H$\alpha$ emission during this
  phase. Within this scheme, the change to a single spectral type at
  all wavelengths and emergence of strong H$\alpha$ emission are
  consistent with the accretion columns becoming optically thin, as
  the accretion rate drops. There is a strong analogy here with the
  dwarf novae class of interacting binaries, which show similar
  behaviour during the decline from outbursts of high mass-transfer
  rate.
  
  The most important consequence of this interpretation is that these
  objects bring into question the association of Weak-Line T Tauri
  stars (WTTs) with non-accreting or discless objects. In light of
  this result we consider the justification for this paradigm.
\end{abstract} 


\section{Introduction}
\label{sec:introduction}
T Tauri stars are divided into classical T Tauri stars (CTTs) and weak
T Tauri stars (WTTs) largely on the basis of their H$\alpha$ emission.
The cutoff is usually taken at 10\AA, with values above this
representing a CTTs \citep{appenzeller89}. It has long been argued that
the difference between CTTs and WTTs is that the former have discs,
and the latter do not. Indeed, there is some support for this
paradigm, with H$\alpha$ equivalent width correlating with other disc
indicators, such as B-V excess, or infrared colour excess
\citep{herbig98,cabrit90}. However, many problems exist with such a
scheme. The cutoff of 10\AA\, is somewhat arbitrary, and is less
sensitive at earlier spectral types \citep{martin97}. Also, recent
studies have discovered WTTs which show infared excesses indicative of
circumstellar discs \citep{hetem02}. Here we present evidence that at
least some WTTs not only possess accretion discs, but are accreting
from them at a rate which is comparable to the most rapidly accreting CTTs.

An optical spectroscopic and photometric survey of a set of 47 stars
associated with {\em Einstein} X-ray sources in the $\rho$ Ophiuchus
cloud \citep[][hereafter BA92]{ba92}, identified 30 pre-main sequence
stars. Four of these stars (ROXs 21, ROXs 47a, ROXs 2 and ROXs 3) were
identified as exhibiting ``composite'' spectra - the spectral type
inferred from a blue spectral region near H$\beta$ was systematically
earlier than that inferred from a red spectral region surrounding the
TiO band between 7050 and 7200\AA. The spectral types inferred for
these stars are listed in Table~\ref{tab:halpha}. We note here that
the composite spectrum effect was reliably detected in the case of
ROXs 21, 47a and 3. The detection of the effect in ROXs 2 was less
secure; although the blue spectral type of K3 was reliable, the lower
resolution spectrum of this object rendered the red spectral type of
M0 uncertain. Our spectrum published here confirms the red spectral
type of BA92, and hence we consider this star as a bona-fide composite
spectrum object. BA92 also found a discrepancy in red and blue
spectral types for ROXs 29, but the difference in spectral type (K4 vs
K6) was barely significant, and we do not consider this star a
bona-fide composite spectrum object.

BA92 put forward two suggestions for the origin of the composite spectra.
\begin{itemize}
\item{The systems are close binaries, with components of different 
temperatures}.
\item{The systems consist of cool stars surrounded by accretion
discs. The blue spectral region is dominated by luminosity either from
the accretion itself, or from hot spots arising where the accretion
stream hits the star}.
\end{itemize}

However, they were unable to determine which, if either, of these
explanations was the case. The binary hypothesis was examined by
\cite{koresko95}, who performed speckle interferometry for ROXs 21 (SR
12), ROXs 47a (Do-Ar 51) and ROXs 3. He found that ROXs 21 and ROXs
47a were close binary stars with brightness ratios at 1.65 $\mu$m of
0.9 and 0.33 respectively. However, he also found that {\em co-eval}
binaries could not be responsible for the composite spectra effect.
The problem arises as the two components must have comparable
luminosities at 5550\AA. The cool component must then be larger than
the hot component, implying that the hot component is substantially
older than the cool component.

Hence, the nature of the objects discovered in BA92 remains a
mystery. In order to resolve this mystery, we revisited the objects,
more than twenty years later, with medium-resolution spectroscopy in the
blue and red regions. The spectra of all four objects show significant
changes since the observations of BA92. All objects now show strong
H$\alpha$ emission and a diminished composite spectrum effect. The
presence of such variability strongly argues against a binary origin
for the phenomenon. Here we suggest that the composite spectrum
phenomenon, and the variability witnessed, is caused by changes in the
accretion state of the young stars.

Section~\ref{sec:obs} describes the observations taken, whilst the
results are shown in section~\ref{sec:results}. In
section~\ref{sec:discussion}, we outline the scheme we propose to
explain these results, and discuss its consequences and in
section~\ref{sec:conclusions} we summarise our conclusions.

\section{Observations}
\label{sec:obs}
On the nights of 2002 Jan 27 and 28 we obtained 5960--7640\AA\, spectra
of ROXs 21, ROXs 47a, ROXs 2 and ROXs 3 using the RGO spectrograph on
the Anglo-Australian Telescope (AAT) at Siding Spring, Australia.  The
1200R grating, blaze-to-camera, in conjunction with the MITTL chip
gave a resolution of 1.5\AA.  The exposure time was 120 secs for each
object. The spectra were extracted using the optimal extraction
algorithms in {\sc figaro} \citep{figaro}.  Observations of the A-type
star SAO184424 were used to correct the spectra for telluric features.
Regular observations of arc and flatfield frames were used to
calibrate the data.

In addition, on the nights of 2002 August 30 and 2002 September 1, we
obtained 3770--4890\AA\, spectra of a sample of 23 of the 30 pre-main
sequence objects oberved by BA92 using the 2dF spectrograph on the AAT
\citep{lewis02}.  Included in the sample were the four composite
spectra objects. Each spectrum consists of 5$\times$300 sec exposures.
The data were taken in service mode by E. Corbett.  The 1200B grating
was used, giving a resolution of 2.2\AA.  Arc and flatfield frames
were obtained after each target exposure, along with offset sky frames
(which may be used to calibrate fibre throughput).  The data were
reduced using the {\sc 2dfdr} data reduction package, with default
settings \citep{lewis02}.

\section{Results}
\label{sec:results}
\begin{figure}
\includegraphics[scale=0.35,angle=-90]{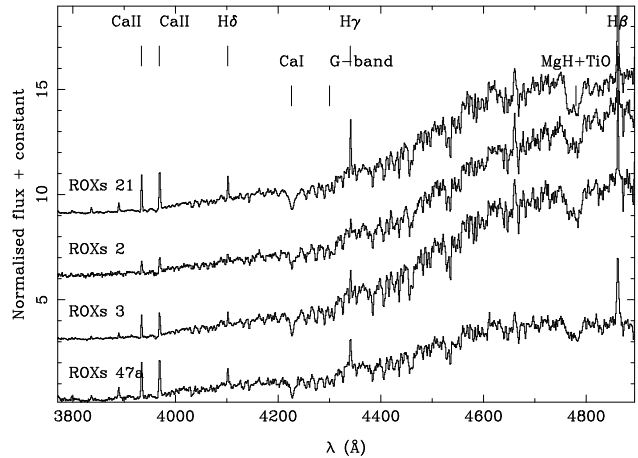}
\caption{2dF blue spectra of the BA92 objects, obtained in August
  2002. The spectra have been normalised and an offset has been
  applied to the y-axis of the spectra for clarity.}
\label{fig:blue}
\end{figure}

\begin{figure}
\includegraphics[scale=0.35]{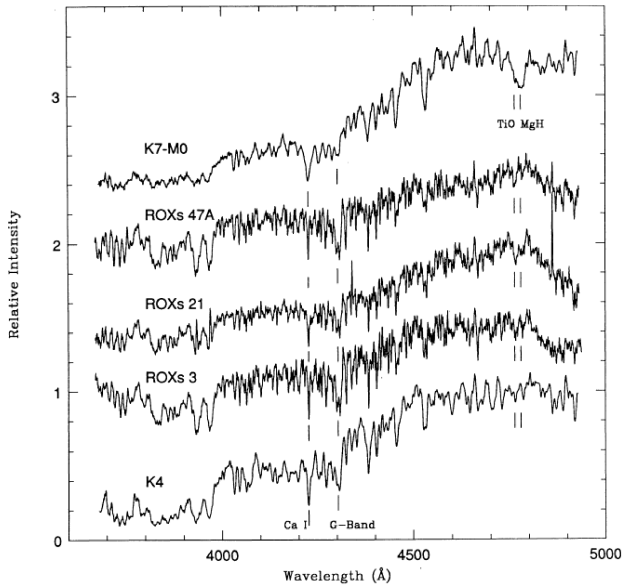}
\caption{The original spectra of the BA92 objects, obtained between
1983 and 1986. The spectra have been normalised, and an offset has
been applied to the y-axis of the spectra for clarity. Also shown is the
spectrum of a K4 dwarf (bottom) and a K7/M0 dwarf (top).}
\label{fig:ba_blue}
\end{figure}

\begin{figure}
\includegraphics[scale=0.35,angle=-90]{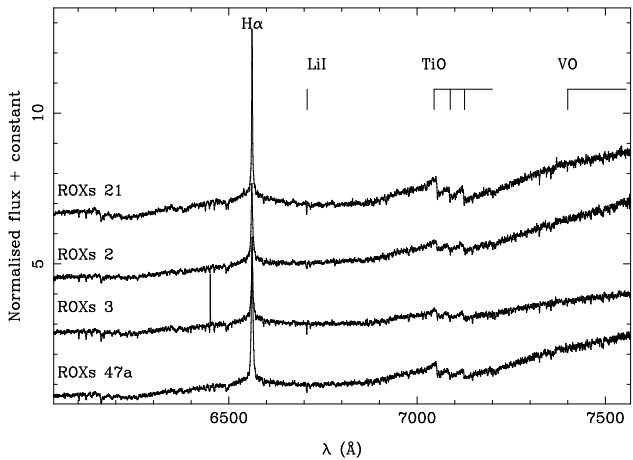}
\caption{RGO red spectra of the BA92 objects, obtained in January
  2002. The spectra have been normalised and an offset has been
  applied to the y-axis of the spectra for clarity.}
\label{fig:red}
\end{figure}

Figure~\ref{fig:blue} shows the 3770--4890\AA\, spectra of ROXs 21,
ROXs 47a, ROXs 2 and ROXs 3. Shown for comparison in
Figure~\ref{fig:ba_blue} are the original data from BA92. The
variability is striking. First, whilst the later spectra of all four
objects are characterised by emission lines of the Balmer series and
CaII, the original data show no Balmer emission, and strong, broad
CaII absorption.  Only one object (ROXs 21) shows any sign of CaII or
Balmer emission in the ealier data.  In all cases, the increase in
emission line strength is significant and striking, even when our data
are degraded to the resolution and signal to noise of the BA92 spectra.
Second, and most intriguing, the prominent G-band seen in the original
data at around 4300\AA\, is much weaker in the current data. The
G-band is visible in dwarfs from mid-F to late-K type, and peaks in
strength around G5V.

With the weakening of the G-band, it is clear that the spectral type
inferred from the blue spectral region is later in our data than in
the data of BA92. In fact, BA92, used the continuum shape and G-band
strength to infer spectral types of K2V--K4V for these objects.
Comparison of our spectra and the K7/M0V spectrum in
Figure~\ref{fig:ba_blue} shows that the spectral type inferred from
the blue spectral region is now closer to late-K or early M.  The
implications of this are discussed in section~\ref{sec:discussion}.

The red spectral region has also seen significant changes since the
original data were taken. BA92 classified all four objects as weak
T-Tauri stars on the basis of their H$\alpha$ emission.
Figure~\ref{fig:red} shows our recent AAT spectra. A glance at this
figure shows that the H$\alpha$ emission properties of the objects has
changed markedly: the red spectra of all four objects are now dominated
by strong H$\alpha$ emission. Table~\ref{tab:halpha} shows the changes
in H$\alpha$ equivalent width between the two observations. In all
cases the equivalent width of H$\alpha$ emission has dramatically
increased. The change is large enough such that all four objects would
now be classified as classical T Tauri stars, on the basis of their
H$\alpha$ emission.  The H$\alpha$ emission is also very broad.
\cite{white03} show that the H$\alpha$ line widths at 10\% of peak
intensity are often a better indicator of ongoing accretion than
H$\alpha$ equivalent width.  They suggest that stars with 10\% widths
of $>270$km\,s$^{-1}$ are accreting, regardless of spectral type. All
four objects show 10\% widths significantly in excess of this value
(see Table~\ref{tab:halpha}). 

Despite the change in H$\alpha$ emission properties, the spectral type
inferred from the red spectral region remains unchanged with respect
to that found by BA92.  This removes any uncertainty that the objects
we observed are the same objects observed by BA92. We now discuss the
implications of these results.

\begin{table}
\caption[]{H$\alpha$ equivalent width variations between this paper
and BA92. Also shown is the H$\alpha$ full-width at 10\% of maximum intensity
for our data, and the spectral types inferred by BA92. The first entry is
the spectral type inferred from the region bluewards of H$\beta$, the
second entry is the spectral type inferred from the region around
H$\alpha$.}
\begin{center}
\begin{tabular}{lcccc}
\multicolumn{1}{c}{Object} & \multicolumn{1}{c}{BA92}
& \multicolumn{1}{c}{This paper} & \multicolumn{1}{c}{This paper} & \multicolumn{1}{c}{BA92}\\
& & & &\\
& \multicolumn{1}{c}{EW (\AA)} & \multicolumn{1}{c}{EW (\AA)}
& \multicolumn{1}{c}{FW10\% (km\,s$^{-1}$)} & \multicolumn{1}{c}{Sp Type}\\
& & & &\\
\hline
& & & &\\
ROXs 21     & 4.5  & 17.6$\pm$0.2 & 370$\pm$30 & K4/M2.5V\\
ROXs 47a    & 9.2  & 24.4$\pm$0.2 & 460$\pm$30 & K2/K7-M0V\\
ROXs 2      & 2.8  & 14.3$\pm$0.2 & 550$\pm$30 & K3/M0V\\
ROXs 3      & 2.3  & 12.2$\pm$0.2 & 500$\pm$30 & K3/M0V\\
& & & &\\
\hline
& & & &\\
\end{tabular}
\end{center}
\label{tab:halpha}
\end{table}

\section{Discussion}
\label{sec:discussion}

\subsection{The nature of composite spectra}
Any theory which seeks to explain the composite spectra observed in these
stars must explain these key facts.
\begin{itemize}
\item{A composite spectrum always shows a hot spectral type in the blue, and
a cooler spectral type in the red.}
\item{The phenomenon is transitory, and decays on a timescales of less than
or of the order 20 years.}
\item{Coupled with the disappearance of the composite spectrum, is the
emergence of an emission line spectrum from CaII and the Balmer lines
in the blue region, and a dramatic rise in the equivalent width of
H$\alpha$ emission in the red region.}
\end{itemize}

Considering these facts, we can reject binarity as a cause for
composite spectra; it is unlikely that the spectral type of a star can
change by several subclasses in $\lta$20 years. A second possibility
is that the composite spectra are caused by cool starspots on the
surface of a hot young star. This, too, seems unlikely. On a
spot-covered star, the overall light is normally dominated by the hot, bright,
spot-free photosphere. BA92 calculated that to produce the composite
spectrum effect observed required 93\% spot coverage.  Given that the
spectrum in the blue region has also now cooled, starspots would now
have to cover the entire star to explain the observed changes.

The most promising scheme is that the composite spectra are caused by
accretion. In this scheme the source for the hot, blue component is
the accretion luminosity itself. Unlike the veiling luminosity
typically seen in CTTs, which is featureless and fills in stellar
absorption lines, we are suggesting that the accretion luminosity,
arising from optically thick accretion columns or in hot spots on the
stellar surface, itself exhibits hot spectral features, like the
G-band at 4300\AA. The stellar spectrum is then completely hidden by
this accretion luminosity.  An optically thick accretion flow would
explain the absence of strong H$\alpha$ emission in this phase. An
attractive property of the accretion model is that the observed
spectral changes are easily understood as a fall in the accretion
rate. First we consider the blue spectral region. As the accretion
rate drops, the accretion flow becomes optically thin, and the
luminosity of the flow reduces drastically. The hot spectral features
will naturally disappear at this point. The drop in accretion
luminosity explains the change to a red continuum, and emergence of an
emission line spectrum as the veiling caused by the luminosity
disappears. In the red spectral region, the observed changes are
simpler, because this region is relatively unaffected by the accretion
emission. In this case, the main observed change is a rise in the
strength of H$\alpha$ emission as the accretion columns become
optically thin.

This behaviour is not without precedent. Dwarf novae are binary stars
consisting of a low-mass red star transferring mass, via an accretion
disc to a white dwarf companion. They exhibit frequent outbursts,
which are linked to an instability in the accretion disc which causes
sudden increases in accretion rate. These objects are characterised in
their quiescent state by a strong emission line spectrum. During
outburst, however, high accretion rates mean the disc atmosphere
becomes optically thick, and an absorption line spectrum, arising from
the disc photosphere, is observed \cite[e.\,g.][]{mansperger90}.  The
similarities between this behaviour and the behaviour seen in the BA92
objects are striking.

The most obvious implication of this explanation is that here are four
stars which may exhibit high accretion rates in the WTT phase. If the
blue luminosity in the BA92 objects is dominated by accretion power,
we can calculate the required accretion rates from the observed B and
V magnitudes, after correction for reddening. We assume the accretion
emission is a black body, with its temperature of 5200K taken to
correspond to the blue spectral type observed by BA92. To reproduce
the composite spectrum effect we scale the luminosity of the accretion
source until it contributes roughly twice the stellar luminosity in
the V-band.  We use the photometry and reddening of BA92, although we
note that the BA92 derived the reddening from the V-R colours of the
objects. Since the intrinsic V-R would have been affected by accretion
luminosity these reddening values are likely to be underestimated,
implying that the accretion rates derived here are a lower limit. In
order to match the observed B and V magnitudes implies the accretion
source has a luminosity of 0.1 L$_{\odot}$. We assume a stellar mass
of 0.5 M$_{\odot}$ and an inner accretion disc radius of 5
R$_{\odot}$, which corresponds to 3 stellar radii at an age of 1 Myr
\citep{dantona97}.  With these values a luminosity of 0.1 L$_{\odot}$
imply accretion rates of at least 10$^{-7}$M$_{\odot}$yr$^{-1}$,
higher than the average value of 10$^{-8}$M$_{\odot}$yr$^{-1}$ found
for CTTs \citep{hartmann98}. Although our value of
10$^{-7}$M$_{\odot}$yr$^{-1}$ is only a rough estimate, it is
certainly true that the composite spectrum objects were accreting at a
rate comparable to that seen in CTTs. This would be a startling
reversal of the paradigm in which strong H$\alpha$ emission is taken
as evidence of strong accretion onto the young star, and weak
H$\alpha$ emission denotes a non-accreting source.

\subsection{Near-infrared excesses}

It is pertinent to consider if there is other evidence for the
existence of accretion discs in these stars. Circumstellar discs can
produce strong infrared emission significantly in excess of the
stellar luminosity. Such excess infrared emission is easily detected
in a near-infrared colour-colour diagram \cite[see][for
example]{haisch01}. We collated JHK photometry for the composite
spectrum stars from a number of sources
\citep{ba92,walter94,rss76,2mass,jensen97}.  The data were transformed
to the {\sc UKIRT} photometric system using the transformations in
\citet{carpenter01} and an average of each magnitude was taken. None
of the four stars showed any evidence for a near infrared excess when
placed on a JHK colour-colour diagram.  L band photometry was also
available for ROXs 47a and 21. JHKL colour-colour diagrams are a much
more sensitive disc indicator than JHK diagrams, often revealing
evidence for circumstellar discs in objects showing no JHK excess
\cite[see][for example]{haisch01}. In this case, ROXs 47a showed a K-L
excess of a few tenths of a magnitude, whilst ROXs 21 showed no K-L
excess. The K-L excess in ROXs 47a may, however, be due to a low-mass
companion \citep{simon87}. Furthermore, we caution about
over-interpretation of these data; T Tauri stars are intrinsically
variable in the near infrared, with variability up to 1 magnitude not
uncommon \citep{eiroa02}. As the photometry obtained above was not
obtained simultaneously it seems plausible that variability could mask
any near infrared excess present.

Although there is a strong correlation between near-infared excess and
other disc indicators, the absence of a near infrared excess does not
necessarily imply the absence of an accretion disc. For example,
\cite{rebull02} found several stars with very strong H$\alpha$
emission, presumably arising in accretion columns, which exhibited no
near infrared excesses. This presumably means it is possible to
maintain significant accretion rates onto a young star through a disc
which is optically thin in its inner parts.

\subsection{Time-dependant behaviour of the composite spectrum state.}

To better understand the physics of what might cause changes in
accretion rate it is pertinent to consider the time-dependance of
these changes. For example, how long does the high accretion state
last? Is the transition between high and low states short?
 
The BA92 sample consists of 30 X-ray selected pre-main sequence stars
which also exceed a given optical flux limit. In their sample, 4 out
of 30 objects show a composite spectrum effect. Naively, this suggests
that $\sim$10\% of T Tauri stars are in such a state (but see below).
Twenty years later, our 2dF spectroscopy shows that all four objects
are now in a low accretion rate state - this argues that the high
state is likely to last less than 20 years, with long-duration high
states of 50 years or more being very unlikely.  Furthermore, in the
study of BA92, the composite spectrum effect was either absent or
showed similar amounts of spectral type mismatch in all objects. In
other words, no objects were observed in a transition between high and
low states. This then implies that the transition time
between the two states is short - statistical arguments suggest the
transition time is unlikely to be longer than 0.5 years.

The timescale of 0.5 years is interesting because it suggests a
physical mechanism for the change in accretion state. In dwarf novae,
the cause of the accretion rate variations is a thermal-viscous
instability (see \citet{lasota01} for a review). A similar thermal-viscous
instability is believed to cause the FU Orionis phenomena in young stars.
In dwarf novae, the accretion rate changes on the thermal timescale
for the disc. The thermal timescale for the inner region of T Tauri
discs is roughly half a year (at a radius of 0.1 AU). This suggests a
thermal-viscous instability in the inner disc may be responsible for
the change in accretion rates between the two states.

Interestingly, none of the nineteen other objects we observed from the
BA92 sample underwent a change from a low to a high state - the
spectral type inferred from our 2dF spectra agreed with the
classification of BA92, with the exception of the four composite
spectrum objects\footnote{ROXs 4 did show a discrepancy with the BA92
spectral type of K4 being incompatible with our classification as
early-M, however close inspection of the spectrum published in BA92
leads us to conclude that this star was mis-classified by the
authors.}. In other words, the sample of BA92 contained four high
state objects and no low state objects. Although we have observed four
high-low transitions, no low-high transitions have occurred within the
BA92 sample. This suggests that the sample of BA92 was strongly
biassed towards objects which were in a high state at the time of
their observations, and that in fact the fraction of objects in a high
state at any given time may be much lower than 10\%.  Only
simultaneous blue/red spectroscopy of an unbiassed sample of pre-main
sequence stars can resolve this question.

\subsection{H$\alpha$ variability in T Tauri stars}

We note here that this is not the first evidence for large changes in
H$\alpha$ emission strength in T Tauri stars. For example,
\cite{joergens01} noted that the equivalent width of H$\alpha$
emission in the T Tauri star RX J1608.6-3922 varied from 7--14\AA.
\cite{herbig98} noted that several of the young stars in the cluster
IC 348 showed variable H$\alpha$ emission, with some objects changing
from 2\AA\, to $>10$\AA\, in equivalent width. In $\lambda$-Ori, the
survey of \cite{duerr82} catalogued 82 stars with H$\alpha$ emission,
which they categorised as ``strong'', ``medium'' and ``weak''.  The
region was revisited by \cite{dolan01}, who found that the H$\alpha$
equivalent widths of the ``strong'' sample included lines as weak as
6\AA, a fact they attribute to variability. Then there is the
remarkable case of KH 15D - a weak T Tauri star which shows eclipses
thought to be caused by inner disc material. During these eclipses the
H$\alpha$ equivalent width rises from around 2\AA\, to nearly
30--50\AA\, \citep{hamilton02}.

Since none of these studies have incorporated spectroscopy in the blue
spectral region, it is not possible to say if these changes in
H$\alpha$ equivalent width are accompanied by the changes we show
here. It is clear, however,  that there are many examples in the
literature where objects have changed their nature from classic to
weak T Tauri stars and vice-versa.

\subsection{The link between H$\alpha$ and discs in T Tauri stars}

It seems certain that strong (EW $> 10$ \AA) H$\alpha$ emission is
indicative of the presence of active accretion and hence, an accretion
disc. Chromospheric emission from H$\alpha$ rarely rises above 10\AA\,
\citep{martin97}, so an additional source of H$\alpha$ emission must
be found.  Furthermore, in stars with H$\alpha$ emission of EW
$>10$\AA, the strength of H$\alpha$ emission is proportional to the
B-V excess \citep[e.g.][]{herbig98}. T Tauri stars with strong
H$\alpha$ also show the infrared excesses thought to arise from the
inner accretion disc \citep{cabrit90}.

However, it seems that the conclusion that WTTs lack discs is not
correct.  In fact, it seems certain that at least some WTTs do possess
discs. For example, in the case of KH 15D, which is a WTTs, it is
material in a circumstellar disc that is believed to be responsible for
eclipses in this system \citep{herbst02}. \cite{weintraub02} detect
molecular hydrogen, believed to originate in an accretion disc, again
around a WTTs.  Also, the variability of H$\alpha$ emission on
timescales as short as twenty years argues against its usefuleness as
a disc indicator (it is unlikely that the discs themselves could be
coming and going on these timescales). Furthermore, in a sample of 27
WTTs, \cite{hetem02} find that 60\% of the objects show infrared
excesses which indicate a circumstellar disc.

In at least some of these cases it is tempting to explain the failure
of the accepted paradigm by stating that WTTs may or may not have
discs, but that these discs are currently not accreting onto the
central star. If our interpretation of the composite spectrum
phenomenon is correct, however, we are forced to accept that not only
are some WTTs accreting, but they are accreting at rates higher than,
or comparable to most CTTs.

The presence of active accretion discs amongst the WTTs population is
an important result. It suggests that there is no fundamental
difference between CTTs and some WTTs, other than the current
accretion rate.  This might explain, amongst other things, why an
evolutionary link between CTTs and WTTs has been difficult to
establish. Furthermore, if the time-averaged accretion rate in
T Tauri stars is higher than previously suspected, this might explain
age spreads seen in some star forming regions. An accretion rate of
10$^{-7}$M$_{\odot}$yr$^{-1}$, as seen in the BA92 objects is high
enough to upset the thermal balance of a T Tauri star, moving it back
towards the birthline in a HR diagram \citep{tout99}.

\section{Conclusions}
\label{sec:conclusions}
We have obtained new spectra of the objects found to exhibit composite
spectra by \cite{ba92}. The composite spectrum effect has diminished
in the intervening years; the blue spectral type now agrees more
closely with that inferred from the red spectral region.
In addition, a series of emission lines has appeared, and the
equivalent width of H$\alpha$ emission has risen drastically. These
spectral changes can be understood if the composite spectrum effect is
a result of the blue spectrum being dominated by hot, optically thick
emission from the accretion flow. The luminosity in the blue then
implies accretion rates of 10$^{-7}$M$_{\odot}$yr$^{-1}$. An important
result of this explanation is that the paradigm relating H$\alpha$
emission to accretion discs is incomplete; at least some WTTs both
posess accretion discs, and are accreting from them at a high rate.

\section*{\sc Acknowledgements}
SPL is supported by PPARC. The Anglo-Australian Telescope is operated
at Siding Spring by the AAO. We thank Elizabeth Corbett for carrying
out the 2dF observations. The authors acknowledge the data analysis
facilities at Exeter provided by the Starlink Project which is run by
CCLRC on behalf of PPARC.

\bibliographystyle{mn2e}
\bibliography{abbrev,refs}

\begin{thebibliography}{}

\bibitem[\protect\citeauthoryear{{Appenzeller} \& {Mundt}}{{Appenzeller} \&
  {Mundt}}{1989}]{appenzeller89}
{Appenzeller} I.,  {Mundt} R.,  1989, \aapr, 1, 291

\bibitem[\protect\citeauthoryear{{Bary}, {Weintraub} \& {Kastner}}{{Bary}
  et~al.}{2002}]{weintraub02}
{Bary} J.~S.,  {Weintraub} D.~A.,    {Kastner} J.~H.,  2002, \apjl, 576, L73

\bibitem[\protect\citeauthoryear{Bouvier \& Appenzeller}{Bouvier \&
  Appenzeller}{1992}]{ba92}
Bouvier J.,  Appenzeller I.,  1992, A\&AS, 92, 516

\bibitem[\protect\citeauthoryear{{Cabrit}, {Edwards}, {Strom} \&
  {Strom}}{{Cabrit} et~al.}{1990}]{cabrit90}
{Cabrit} S.,  {Edwards} S.,  {Strom} S.~E.,    {Strom} K.~M.,  1990, \apj, 354,
  687

\bibitem[\protect\citeauthoryear{{Carpenter}}{{Carpenter}}{2001}]{carpenter01}
{Carpenter} J.~M.,  2001, AJ, 121, 2851

\bibitem[\protect\citeauthoryear{{Cohen}}{{Cohen}}{1988}]{figaro}
{Cohen} J.~G.,  1988, in Robinson L.~B.,  ed., Instrumentation for Ground-Based
  Optical Astronomy, Present and Future. {The FIGARO Package for Astronomical
  Data Analysis}.
Springer-Verlag, New York, p.~448

\bibitem[\protect\citeauthoryear{{Cutri}, {Skrutskie}, {van Dyk}, {Beichman},
  {Carpenter}, {Chester}, {Cambresy}, {Evans}, {Fowler}, {Gizis} \& {et.
  al.}}{{Cutri} et~al.}{2003}]{2mass}
{Cutri} R.~M.,  {Skrutskie} M.~F.,  {van Dyk} S.,  {Beichman} C.~A.,
  {Carpenter} J.~M.,  {Chester} T.,  {Cambresy} L.,  {Evans} T.,  {Fowler} J.,
  {Gizis} J.,    {et. al.} 2003, VizieR Online Data Catalog, 2246, 0

\bibitem[\protect\citeauthoryear{{D'Antona} \& Mazzitelli}{{D'Antona} \&
  Mazzitelli}{1997}]{dantona97}
{D'Antona} F.,  Mazzitelli I.,  1997, Mem. Soc. Astr. It., 68, 807

\bibitem[\protect\citeauthoryear{{Dolan} \& {Mathieu}}{{Dolan} \&
  {Mathieu}}{2001}]{dolan01}
{Dolan} C.~J.,  {Mathieu} R.~D.,  2001, \aj, 121, 2124

\bibitem[\protect\citeauthoryear{{Duerr}, {Imhoff} \& {Lada}}{{Duerr}
  et~al.}{1982}]{duerr82}
{Duerr} R.,  {Imhoff} C.~L.,    {Lada} C.~J.,  1982, \apj, 261, 135

\bibitem[\protect\citeauthoryear{{Eiroa}, {Oudmaijer}, {Davies}, {de Winter},
  {Garz{\' o}n}, {Palacios}, {Alberdi}, {Ferlet}, {Grady}, {Cameron}, {Deeg} \&
  {et. al.}}{{Eiroa} et~al.}{2002}]{eiroa02}
{Eiroa} C.,  {Oudmaijer} R.~D.,  {Davies} J.~K.,  {de Winter} D.,  {Garz{\'
  o}n} F.,  {Palacios} J.,  {Alberdi} A.,  {Ferlet} R.,  {Grady} C.~A.,
  {Cameron} A.,  {Deeg} H.~J.,    {et. al.} 2002, \aap, 384, 1038

\bibitem[\protect\citeauthoryear{{Gregorio-Hetem} \& {Hetem}}{{Gregorio-Hetem}
  \& {Hetem}}{2002}]{hetem02}
{Gregorio-Hetem} J.,  {Hetem} A.,  2002, \mnras, 336, 197

\bibitem[\protect\citeauthoryear{{Haisch}, {Lada} \& {Lada}}{{Haisch}
  et~al.}{2001}]{haisch01}
{Haisch} K.~E.,  {Lada} E.~A.,    {Lada} C.~J.,  2001, \aj, 121, 2065

\bibitem[\protect\citeauthoryear{{Hamilton}, {Herbst}, {Bailer-Jones} \&
  {Mundt}}{{Hamilton} et~al.}{2002}]{hamilton02}
{Hamilton} C.~M.,  {Herbst} W.,  {Bailer-Jones} C.~A.~L.,    {Mundt} R.,  2002,
  AAS, 201, 1

\bibitem[\protect\citeauthoryear{{Hartmann}}{{Hartmann}}{1998}]{hartmann98}
{Hartmann} L.,  1998, {Accretion processes in star formation}.
Cambridge University Press, Cambridge

\bibitem[\protect\citeauthoryear{{Herbig}}{{Herbig}}{1998}]{herbig98}
{Herbig} G.~H.,  1998, \apj, 497, 736

\bibitem[\protect\citeauthoryear{{Herbst}, {Hamilton}, {Vrba}, {Ibrahimov},
  {Bailer-Jones}, {Mundt}, {Lamm}, {Mazeh}, {Webster}, {Haisch}, {Williams},
  {Rhodes}, {Balonek}, {Scholz} \& {Riffeser}}{{Herbst}
  et~al.}{2002}]{herbst02}
{Herbst} W.,  {Hamilton} C.~M.,  {Vrba} F.~J.,  {Ibrahimov} M.~A.,
  {Bailer-Jones} C.~A.~L.,  {Mundt} R.,  {Lamm} M.,  {Mazeh} T.,  {Webster}
  Z.~T.,  {Haisch} K.~E.,  {Williams} E.~C.,  {Rhodes} A.~H.,  {Balonek} T.~J.,
   {Scholz} A.,    {Riffeser} A.,  2002, \pasp, 114, 1167

\bibitem[\protect\citeauthoryear{{Jensen} \& {Mathieu}}{{Jensen} \&
  {Mathieu}}{1997}]{jensen97}
{Jensen} E.~L.~N.,  {Mathieu} R.~D.,  1997, \aj, 114, 301

\bibitem[\protect\citeauthoryear{{Joergens}, {Guenther}, {Neuh{\" a}user},
  {Fern{\' a}ndez} \& {Vijapurkar}}{{Joergens} et~al.}{2001}]{joergens01}
{Joergens} V.,  {Guenther} E.,  {Neuh{\" a}user} R.,  {Fern{\' a}ndez} M.,
  {Vijapurkar} J.,  2001, \aap, 373, 966

\bibitem[\protect\citeauthoryear{Koresko}{Koresko}{1995}]{koresko95}
Koresko C.~D.,  1995, ApJ, 440, 764

\bibitem[\protect\citeauthoryear{Lasota}{Lasota}{2001}]{lasota01}
Lasota J.-P.,  2001, New Astronomy Review, 45, 449

\bibitem[\protect\citeauthoryear{{Lewis}, {Cannon}, {Taylor}, {Glazebrook},
  {Bailey}, {Baldry}, {Barton}, {Bridges}, {Dalton}, {Farrell} \&
  {Gray}}{{Lewis} et~al.}{2002}]{lewis02}
{Lewis} I.~J.,  {Cannon} R.~D.,  {Taylor} K.,  {Glazebrook} K.,  {Bailey}
  J.~A.,  {Baldry} I.~K.,  {Barton} J.~R.,  {Bridges} T.~J.,  {Dalton} G.~B.,
  {Farrell} T.~J.,    {Gray} P.~M.,  2002, \mnras, 333, 279

\bibitem[\protect\citeauthoryear{{Mansperger} \& {Kaitchuck}}{{Mansperger} \&
  {Kaitchuck}}{1990}]{mansperger90}
{Mansperger} C.~S.,  {Kaitchuck} R.~H.,  1990, \apj, 358, 268

\bibitem[\protect\citeauthoryear{{Mart\'{\i}n}}{{Mart\'{\i}n}}{1997}]{martin97}
{Mart\'{\i}n} E.~L.,  1997, \aap, 321, 492

\bibitem[\protect\citeauthoryear{{Rebull}, {Makidon}, {Strom}, {Hillenbrand},
  {Birmingham}, {Patten}, {Jones}, {Yagi} \& {Adams}}{{Rebull}
  et~al.}{2002}]{rebull02}
{Rebull} L.~M.,  {Makidon} R.~B.,  {Strom} S.~E.,  {Hillenbrand} L.~A.,
  {Birmingham} A.,  {Patten} B.~M.,  {Jones} B.~F.,  {Yagi} H.,    {Adams}
  M.~T.,  2002, \aj, 123, 1528

\bibitem[\protect\citeauthoryear{{Rydgren}, {Strom} \& {Strom}}{{Rydgren}
  et~al.}{1976}]{rss76}
{Rydgren} A.~E.,  {Strom} S.~E.,    {Strom} K.~M.,  1976, \apjs, 30, 307

\bibitem[\protect\citeauthoryear{Simon, Howell, Longmore, Wilking, Peterson \&
  Chen}{Simon et~al.}{1987}]{simon87}
Simon M.,  Howell R.~R.,  Longmore A.~J.,  Wilking B.~A.,  Peterson D.~M.,
  Chen W.-P.,  1987, ApJ, 320, 344

\bibitem[\protect\citeauthoryear{{Tout}, {Livio} \& {Bonnell}}{{Tout}
  et~al.}{1999}]{tout99}
{Tout} C.~A.,  {Livio} M.,    {Bonnell} I.~A.,  1999, \mnras, 310, 360

\bibitem[\protect\citeauthoryear{{Walter}, {Vrba}, {Mathieu}, {Brown} \&
  {Myers}}{{Walter} et~al.}{1994}]{walter94}
{Walter} F.~M.,  {Vrba} F.~J.,  {Mathieu} R.~D.,  {Brown} A.,    {Myers} P.~C.,
   1994, \aj, 107, 692

\bibitem[\protect\citeauthoryear{{White} \& {Basri}}{{White} \&
  {Basri}}{2003}]{white03}
{White} R.~J.,  {Basri} G.,  2003, \apj, 582, 1109

\end{thebibliography}

\end{document}